\documentclass[12pt]{iopart}
\usepackage{graphicx}
\usepackage{cite}
\usepackage{amsfonts}
\usepackage{array}
\usepackage{booktabs}
\usepackage{srcltx}

\def\zbar{{\overline{z}}} 
\def\deltabar{{\overline{\delta}}}
\def\nuparbar{{\overline{\nu_\parallel}}}
\def\lamstar{{\lambda_*}}
\begin{document}
\title[Multicritical behavior of the diluted contact process]
      {Multicritical behavior of the diluted contact process}
\author{Silvio R.Dahmen$^{1,3}$, 
Lionel Sittler$^2$, 
and Haye Hinrichsen$^1$}

\address{$^1$ Universit\"at W\"urzburg, Fakult\"at f\"ur Physik und Astronomie,\\
	 Am Hubland, D-97074 W\"urzburg, Germany\\}
\address{$^2$ Universit\"at Duisburg-Essen, Fachbereich Physik, \\ 
	Lotharstra{\ss}e 1, D-47057 Duisburg, Germany}
\address{$^3$ Instituto de F\'{\i}sica da UFRGS\\
	Av. Bento Gon{\c c}alves 9500, 91501-970 Porto Alegre RS, Brazil}

\begin{abstract}
We study a contact process on a two-dimensional square lattice which is diluted by randomly removing bonds with probability $p$. For $p<1/2$ and varying birth rate~$\lambda$ the model was shown to exhibit a continuous phase transition which belongs to the universality class of strongly disordered directed percolation. The phase transition line terminates in a multicritical point at $p=1/2$ and $\lambda=\lambda*=3.55(1)$, where the model can be interpreted as a critical directed percolation process running on a critical isotropic percolation cluster. In the present work we study the multicritical point and its neighboorhood by numerical simulations, discussing possible scaling forms which could describe the critical behavior at the transition.
\end{abstract}

\submitto{Journal of Statistical Mechanics: Theory and Experiment}
\parskip 2mm 
\vglue 5mm

\section{Introduction}

\noindent
Phase transitions from fluctuating phases into so-called absorbing states continue to be an important topic of non-equilibrium statistical physics. A prototypical system that exhibits such a non-equilibrium phase transition is the contact process. It is defined on a $d$-dimensional hypercubic lattice with sites that can be either empty (inactive) or occupied by a particle (active). As time evolves, particles are spontaneously removed at unit rate and created at empty sites at rate $\lambda n/(2d)$, where the so-called birth rate $\lambda$ is a control parameter and $n$ denotes the actual number of occupied nearest neighbor sites. In the inactive phase, where $\lambda$ is small, particle removal dominates and the system approaches a so-called absorbing state without particles from where it cannot escape. However, if $\lambda$ exceeds a certain threshold $\lambda_c$, a steady state with a finite density of particles exists on the infinite lattice and the system is said to be in the active phase. The two phases are separated by a nonequilibrium phase transition at the critical threshold $\lambda=\lambda_c$. This transition is known to belong to the universality class of directed percolation (DP)~\cite{MarroDickman,HinrichsenReview,OdorReview,LubeckReview}.

In spite of the progress in the theory of absorbing phase transitions, there is so far no experiment that reproduces the critical behavior of the transition in a reliable way ~\cite{BrazilReview}. Although various experiments show signatures of a continuous transition on a qualitative level, none of them was able to reproduce the full set of DP critical exponents quantitatively. Designing and performing such an experiment remains one of the most challenging open problems of non-equilibrium statistical mechanics.

There are various reasons why experiments of DP are so difficult to perform. For example, in the contact process the absorbing state is perfectly stable while in realistic situations there is always a small but finite rate for spontaneous creation of activity that washes out the transition. Another possible source of difficulties  in many experiments is the presence of \textit{spatially quenched disorder} due to inhomogeneities of the system. In the contact process this type of disorder can be introduced by initially varying the birth rate~$\lambda$ from site to site and keeping these variations fixed as time evolves. From the field-theoretic point of view, spatially quenched disorder is a marginal perturbation under renormalization group transformations which entirely changes the critical behavior at the transition~\cite{Janssen97}. In fact, numerical studies by Moreira and Dickman~\cite{Moreira96,Dickman98} showed that the critical dynamics of the contact process with quenched disorder evolves on a logarithmic time scale, e.g., the density of particles decays as $\rho(t) \sim [\ln(t)]^{-\deltabar}$ with some exponent $\deltabar$. More recently, Hooyberghs \textit{et al.}~\cite{Hooyberghs03,Hooyberghs04} argued that this type of critical behavior can be related to the random-field Ising model provided that the influence of quenched disorder is strong enough.

\begin{figure}[t]
\begin{center}
\includegraphics[width=80mm]{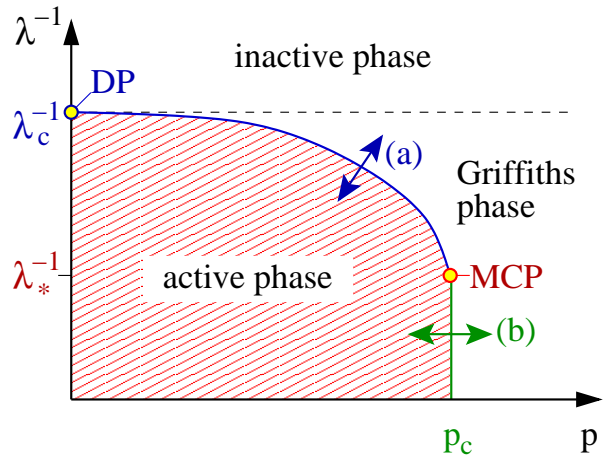}
\caption{
\label{fig:phasediag}
Schematic phase diagram of the diluted contact process as a function of the dilution parameter $p$  and the birth rate $\lambda$. The phase transition lines (a) and (b) are explained in the text. Both lines meet in the multicritical point (MCP) located at the percolation threshold $p=p_c$ and the critical birth rate $\lambda=\lamstar$. The present study focuses on the scaling behavior at and in the vicinity of the multicritical point.
}
\end{center}
\end{figure}

A very simple model, where the effect of disorder can be studied systematically, is the diluted contact process introduced by Noest {\it et al.}~\cite{Noest86,Noest88}. In this model a two-dimensional square lattice is first diluted by randomly removing lattice sites with probability $p$. After diluting the lattice, a contact process runs on the remaining sites according to the usual dynamical rules.  The schematic phase diagram of this model is shown in Fig.~\ref{fig:phasediag}. For $p=0$ (no dilution) the model reduces to the ordinary two-dimensional contact process with a DP phase transition at $\lambda_c = 1.64877(3)$~\cite{Dickman99}. For weak dilution below the percolation threshold of the lattice, $p < p_c$,
the active phase of the model survives, but the critical birth rate $\lambda$ increases with $p$ in order to compensate the missing neighbors. For strong dilution beyond the percolation threshold, $p>p_c$, the lattice decomposes into many disconnected finite clusters so that a steady state with a positive density of active sites cannot exist. Right at the percolation threshold $p=p_c$, the active phase survives only on the infinite percolation cluster provided that the birth rate $\lambda$ is larger than some threshold $\lamstar$. Thus, the active phase is bounded by two different phase transition lines, a curved and a vertical one, denoted by (a) and (b) in the figure. The two lines meet in a multicritical point (MCP). 

Although the scaling properties along these transition lines (a) and (b) are already well understood, the scaling behavior \textit{at} the multicritical point (MCP) has not been studied so far. At the multicritical point the model can be viewed as a critical DP process running on top of critical frozen clusters of the diluted lattice at the percolation threshold. As both models -- isotropic percolation providing the support and the DP process running on top of it -- are scale-free at the multicritical point and characterized by different critical properties, one expects a non-trivial interplay of both universality classes leading to an unconventional scaling behavior. The purpose of this paper is to present numerical results and to test possible scaling forms that could describe the critical behavior of the diluted contact process at and in the vicinity of the multicritical point. To this end the paper starts with a short review of various known types of scaling. In Sect.~\ref{NumericsSect} we present our numerical results at and close to the multicritical point, testing several possible types of scaling. These results are then interpreted in the conclusions in Sect.~\ref{ConclusionSect}.

\section{Conventional and non-conventional scaling forms}
\label{ScalingSect}

In this section we briefly review several types of phenomenological scaling laws which are associated with different parts of the phase diagram in Fig.~\ref{fig:phasediag}. A more comprehensive discussion can be found in Ref.~\cite{Hooyberghs04}. 

\subsection{Conventional scaling}

The \textit{clean} contact process without dilution ($p=0$) is known to exhibit ordinary power law scaling. This well-established type of scaling assumes that all relevant quantities such as distances, time intervals, the system size as well as order- and control parameters can be multiplied by a factor $\Lambda$ raised to some power in such a way that the long-range properties of the system close to the transition remain invariant. In the contact process the order parameter is the density of active sites $\rho(\Delta,t,L)$ which depends on the time parameter $t$, the distance from the critical point $\Delta=\lambda-\lambda_c$, and the lateral system size $L$. In the vicinity of the clean transition (marked by 'DP' in the phase diagram), where $\Delta$ is small, this order parameter scales under changing lengths by a factor $\Lambda$ as
\begin{equation}
\label{ConventionalRescaling}
\rho (\Delta,t,L) = \Lambda^{x}  \, \rho \Bigl(\Lambda^{-1/\nu_\perp}\Delta, \Lambda^z t,\Lambda L\Bigr)\,,
\end{equation}
where $x= \beta/\nu_{\perp}$ and $z=\nu_\parallel/\nu_\perp$ are certain critical exponents. Scaling forms can be obtained by choosing $\Lambda$ in such a way that one of the arguments on the r.h.s. becomes constant. For example, choosing $\Lambda=t^{-1/z}$ one gets
\begin{equation}
\label{ConventionalScalingForm}
\rho (\Delta,t,L) = t^{-\delta}  \, f\bigl(t^{1/\nu_\parallel}\Delta, t^{-1/z} L\bigr)\,,
\end{equation}
where $\delta=x/z=\beta/\nu_\parallel$ and $f$ is a scaling function. In most cases this scaling function is universal, i.e., its functional form is fully given by the universality class to which the phase transition belongs.

The above scaling form implies that the density of active sites measured in an infinite system at criticality decays as $\rho(t) \sim t^{-\delta}$. Moreover, approaching the transition the correlation length and the correlation time both diverge algebraically as 
\begin{equation}
\xi_\perp \sim \Delta^{-\nu_\perp}\,, \qquad \xi_\parallel \sim \Delta^{-\nu_\parallel}\,,
\end{equation}
meaning that both scales are related by $\xi_\parallel \sim \xi_\perp^z$.

\subsection{Activated scaling}

Along the transition line (a) in Fig.~\ref{fig:phasediag} the diluted contact process is known to display an unconventional type of scaling behavior that has been termed as \textit{activated scaling} or \textit{strong disorder scaling} in the literature. This type of scaling is characterized by an ultra-slow temporal evolution which depends on $\ln t$ instead of $t$. Activated scaling may be interpreted as a limiting case of conventional scaling when the exponent $\nu_\parallel$ is taken to infinity. The resulting scaling laws are very similar to the conventional ones, the only difference being that $t$ is replaced by $\ln t$ and that the exponents may have different values. 

In the case of activated scaling the characteristic length scale still diverges algebraically whereas the correlation time diverges \textit{exponentially} as the critical point is approached:
\begin{equation}
\xi_\perp \sim \Delta^{-\nu_\perp}\,, \qquad \xi_\parallel \sim \exp\left(a\Delta^{-\nuparbar}\right)\,.
\end{equation}
Here $a$ is a constant and $\nuparbar$ is an exponent analogous to $\nu_\parallel$ in the conventional case. This means that time and length scales are no longer related by $\xi_\parallel=\xi_\perp^z$, instead  $\xi_\perp$ is related to the \textit{logarithm} of  $\xi_\parallel$ through
\begin{equation}
\ln \xi_{\parallel} \sim {\xi_{\perp}}^{\zbar}\,,
\end{equation}
where $\zbar=\nuparbar/\nu_\perp$.\footnote{In previous works the exponent $\zbar$ is also denoted as $\psi$.} The value of the exponents $\nuparbar$ and $\zbar$ may depend on the strength of disorder, i.e., the site dilution probability $p$. 

In the case of activated scaling the usual relation for the density $\rho$ upon rescaling $L\rightarrow \Lambda L$ in Eq.~(\ref{ConventionalRescaling}) is replaced by
\begin{equation}
\label{ActivatedScaling}
\rho (\Delta,t,L) = \Lambda^{x} \, \rho \Bigl(\Lambda^{-1/\nu_\perp}\Delta, \,\Lambda^{\zbar} [\ln (t/t_0)],\,\Lambda L\Bigr)\,,
\end{equation}
where $x=\beta/\nu_\perp$ and $\Delta$ denotes the distance from the transition line. Choosing $\Lambda=[\ln (t/t_0)]^{-1/\zbar}$ one gets
\begin{equation}
\label{ActivatedScalingForm}
\rho (\Delta,t,L) = [\ln (t/t_0)]^{-\overline{\delta}}  \, F\Bigl([\ln (t/t_0)]^{1/\nuparbar}\Delta,\,, [\ln (t/t_0)]^{-1/\zbar} L\Bigr)\,,
\end{equation}
where $F$ is a scaling function and $\overline{\delta}=x/\zbar=\beta/\nuparbar$ plays a role of a decay exponent analogous to $\delta$ for conventional scaling. Note that whenever one takes the logarithm of a parameter, it has to be divided by a constant to make the argument dimensionless. For example, in the scaling form given above, $t_0$ plays the role of an elementary time scale. In a numerical simulation such a constant is expected to be of order $1$ in natural units of Monte Carlo sweeps.

As shown in a recent paper by Vojta and Lee~\cite{VojtaLee06}, activated scaling is not only valid along line (a) but also along the vertical transition line denoted by (b) in Fig.\ref{fig:phasediag}. Here the supporting lattice is at the percolation threshold and thus decomposes into many clusters according to a scale-free distribution. Each of these clusters hosts an independent contact process. In the limit $t \to \infty$ a nonzero steady-state density can only survive on the infinite percolation cluster while finite clusters enter the absorbing state by rare fluctuations and thus do not contribute asymptotically. 

Since the infinite cluster has a fractal dimension $D_f<2$ and therefore covers only a vanishing fraction of lattice sites  in an infinite system, it does not contribute to the density of active sites $\rho(t)$. This means that $\rho(t)$ is dominated by contact processes on finite clusters and hence approaches zero as $t$ goes to infinity. However, this decay is extremely slow since contact processes on larger clusters may have a very long life time. Vojta and Lee showed that the transition along line (b) obeys activated scaling, i.e., we can use the same scaling form as for strongly disordered DP, although with a different set of exponents.

Vojta and Lee also proved that the exponents $\beta$ and $\nu_\perp$ on line (b) coincide with the usual order parameter and correlation length exponents $\beta_c=5/36$ and $\nu_c=4/3$ of isotropic percolation in two dimensions, i.e., all static features are dictated by the scaling properties of the supporting lattice at the percolation threshold. For example, in the vicinity of the critical line the steady-state density $\rho^{stat}$ and the correlation length $\xi_\perp$ are scale as
\begin{equation}
\rho^{stat} \sim \vert p-p_c \vert^{\beta_c} \,, \qquad
\xi_\perp \sim \vert p-p_c \vert^{-\nu_c}\,. \qquad (p<p_c)
\end{equation}
Moreover, they found that the density of active sites decays as
\begin{equation}
\rho(t) \sim [\ln(t/t_0)]^{-\deltabar}\,,
\end{equation}
where $\deltabar = \beta_c/(\nu_c D_f) = 5/91$ is again related to the exactly known static exponents and the fractal dimension of isotropic percolation in two dimensions. In the vicinity of the line (b) sufficiently far away from the multicritical point one finds a Griffiths-like power law decay 
\begin{equation}
\rho(t) \sim (t/t_0)^{-d/z'}
\end{equation}
for $p>p_c$ and a stretched exponential of the form
\begin{equation}
\rho(t) - \rho_{\rm st} \sim \exp\left[\left(\frac{d}{z''}\ln\left(\frac{t}{t_0}\right)\right)^{1-1/d}\right]
\end{equation}
for $p<p_c$, where $\rho_{\rm st}$ again denotes the stationary density. Here $d$ is the spatial dimension while $z'$ and $z''$ are nonuniversal exponents with diverge as $p\to p_c$.

To summarize, both transition lines (a) and (b) obey activated scaling, although with different sets of critical exponents, the latter being related to the exactly known exponents of isotropic percolation in two dimensions. However, the physical mechanisms leading to a logarithmic time dependence are very different in both cases. As pointed out by Vojta and Lee, this difference is also reflected in the early time behavior of the model. To this end one considers the dynamics of a contact process starting from a single active site on a diluted lattice. On line (b), where the contact process is locally supercritical, the cloud of active sites first grows linearly with time until it covers its supporting cluster completely. This cloud then remains in a metastable state until it reaches the absorbing state by a rare fluctuation. Therefore one observes a very fast short-time behavior followed by a logarithmically slow decay. On line (a), however, the observed initial spreading is much slower.

\subsection{Scaling in the Griffiths phase}

Quenched disorder by dilution leads to the emergence of a so-called Griffiths phase which in the present model is located between the transition line and the horizontal line starting at the clean critical point (the dashed line in Fig.\ref{fig:phasediag}). A Griffiths phase is characterized by an algebraic decay of the order parameter with non-universal continuously varying exponents while spatial correlations remain short-ranged. 

The algebraic decay can be understood as follows. Consider a compact finite region of the lattice and let $k$ be the corresponding number of bonds. The probability that none of these bonds was removed during the dilution procedure is $(1-p)^k$, hence the probability to find a fully connected region decays exponentially with its size. Since for large regions with typically more than $-\ln(1-p)$ bonds this probability is extremely small, such fully connected islands have been named as \textit{rare regions} in the literature. For $\lambda>\lambda_c$ each rare region supports a locally supercritical contact process that survives in a metastable active state for a typical time $\tau$ which grows as $\tau \propto \exp(+\alpha k)$ with $k$, where $\alpha>0$ depends on $\lambda-\lambda_c$. The exponentially diverging life time combined with an exponentially decaying likelihood for rare regions is the origin of the observed non-universal power-law behavior in the Griffiths phase.

To see this note that at a given time $t$ only those rare regions are still in the metastable active state whose number of bonds $k$ is typically larger than $\alpha \ln t$. Summing up their contributions to the particle density weighted by the probability $(1-p)^k$ to find such regions, one arrives at an expression proportional to
\begin{equation}
(1-p)^{(\ln t)/\alpha} \;=\; t^{\, \ln(1-p)/\alpha} \,.
\end{equation}
Obviously, the continuously varying exponent $\ln(1-p)/\alpha$ depends on both $p$ and $\lambda-\lambda_c$. 

\subsection{Scaling forms based on continuously varying exponents}
\label{ContExpSubsect}

In the present work we also test another type of logarithmic scaling which has been discussed in the context of multiscaling. The theoretical background for this generalized scaling theory is described in Ref.~\cite{SittlerHinrichsen02} and can be sketched as follows. 

Starting point is to generalize conventional scale transformations
\begin{equation}
\tau(a): 
\quad \left\{
\Delta \to \Lambda^{-1/\nu_\perp} \Delta \,, \quad 
t \to \Lambda^z t \,, \quad 
L \to \Lambda L \,, \quad 
\rho \to \Lambda^{-\beta/\nu_\perp} \rho 
\right\}
\end{equation}
in such a way that the critical exponents $\beta,\nu_\perp,\nu_\parallel$ are replaced by \textit{continuously varying} exponents which may depend on the parameters $\Delta,t$ and $L$. However, the functional form of these exponents is not arbitrary, rather it is restricted by the postulate that $\tau(a)$ can still be interpreted as a scale transformation (dilatation) controlled by a scale parameter $\Lambda$. More specifically, it is required that two subsequent scale transformations by $\Lambda_1$ and $\Lambda_2$ are equivalent to a single transformation by the factor $\Lambda_1\Lambda_2$, i.e., scale transformations should obey the group homomorphism
\begin{equation}
\label{Homomorphism}
\tau(\Lambda_1) \circ \tau(\Lambda_2) = \tau(\Lambda_1\Lambda_2)\,.
\end{equation}
As shown in Ref.~\cite{SittlerHinrichsen02}, this requirement leads to the partial differential equations
\begin{equation}
D[\beta(\Delta,t,L)] = D[\nu_\perp(\Delta,t,L)] = D[\nu_\parallel(\Delta,t,L)]=0
\end{equation}
with the differential operator
\begin{equation}
\label{Generator}
D \;=\; \beta(\Delta,t,L) \, \Delta \frac{\partial} {\partial \Delta} \,+\, \nu_\parallel(\Delta,t,L) \, t \frac{\partial}{\partial t} \,+\, \nu_\perp(\Delta,t,L) \, L \frac{\partial} {\partial L}\,.
\end{equation}
These differential equations constrain the possible functional form of the continuously varying critical exponents. Since $D$ is the generator of generalized scale transformations, these partial differential equations tell us that the exponents may vary with the parameters but only in such a way that they are invariant under scale transformations~$\tau$. Clearly, conventional scaling with constant critical exponents is a possible solution. However, as shown in Ref.~\cite{SittlerHinrichsen02}, there are also other non-trivial solutions. For simplicity let us consider the critical case $\Delta=0$, where $\rho(t,L)$ is a function of only two parameters. One possible solution assumes that the exponents and the density are functions of the quotient
\begin{equation}
\alpha=\frac{\ln (t/t_0)}{\ln (L/L_0)}
\end{equation}
and that the exponents $\nu_\parallel(\alpha)$ and $\nu_\perp(\alpha)$ are related by
\begin{equation}
\label{SittlerRelation}
\nu_\parallel(\alpha)  = \alpha \nu_\perp(\alpha).
\end{equation}
In order to verify this solution, note that Eq.~(\ref{SittlerRelation}) turns the generator $D$ into
\begin{equation}
D=\nu_\parallel(\alpha) \left( \frac{\partial}{\partial \ln t} + \alpha \frac{\partial}{\partial \ln L} \right)
\end{equation}
so that $Dh(\alpha)=0$ for any function $h$ that depends exclusively on $\alpha$. 

In addition, the functional form of the order parameter $\rho(t,L)$ is restricted by the differential equation $D \rho(t,L) = \beta(\alpha)/\nu_\perp(\alpha)$. Solving these differential equations one arrives at the scaling form
\begin{equation}
\ln[\rho(t,L)] \;=\; \ln(L/L_0) G \left( \frac{\ln(t/t_0)}{\ln(L/L_0)} \right) \,+\,
 H \left( \frac{\ln(t/t_0)}{\ln(L/L_0)} \right)
\end{equation}
which can be tested numerically by an appropriate data collapse. Again we divided all relevant quantities by constants to make the arguments of the logarithm dimensionless. Setting $H=0$ this logarithmic scaling form was successfully applied to several systems that are believed to exhibit multiscaling, including self-organized critical sandpile models~\cite{Kadanoff89,Tebaldi99}, DLA-related growth processes~\cite{Wu90,Hinnemann01} and experiments in turbulence~\cite{Meakin86}. However, in spite of its apparent generality the logarithmic scaling form seems to fail in the present case, as will be shown.

\section{Numerical results}
\label{NumericsSect}

\subsection{Definition of the model}
\label{ModelSubSect}

\begin{figure}[t]
\begin{flushright}
\includegraphics[width=140mm]{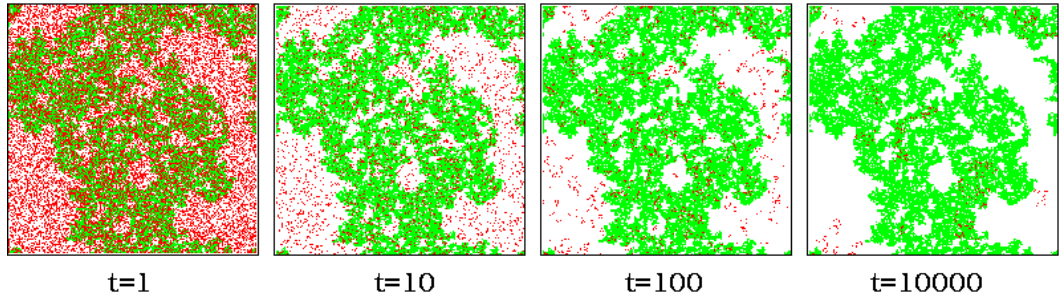}
\end{flushright}
\begin{center}
\caption{
\label{fig:snapshots}
Snapshots of the bond-diluted contact process on a $300 \times 300$ lattice with periodic boundary conditions at the multicritical point. Active sites are represented by red pixels. The maximal cluster, which in a numerical simulation approximates the infinite one, is highlighted by green pixels.
}
\end{center}
\end{figure}

The diluted contact process studied in this paper is defined on a two-dimensional square lattice with periodic boundary conditions and $N=L\times L$ sites which can be either active or inactive. In contrast to Refs.~\cite{Noest86,Noest88,Vojta05}, we will consider a \textit{bond-diluted} contact process, i.e., bonds instead of sites are randomly removed with probability $p$. This type of dilution has the advantage that the percolation threshold of the lattice $p_c=1/2$ is known exactly~\cite{StaufferAharony92}. 

After removing the bonds, all lattice sites are initially occupied by a particle. The model then evolves by random-sequential updates in the same way as the ordinary contact process: For each attempted update one of the particles is randomly selected. With probability $\frac{1}{1+\lambda}$ this particle is removed, otherwise one of the four nearest neighboring sites is randomly selected and a new particle is created provided that the target site is empty and the corresponding bond has not been removed during the initial dilution procedure. Each update attempt corresponds to an average time increment of $\frac{1}{n(1+\lambda)}$, where $n$ is the actual number of active sites. Typical snapshots of a simulation of a single run are shown in Fig.~\ref{fig:snapshots}. The quantities to be measured are averaged over a large number of runs, each of them with an independent realization of quenched randomness.  

As usual, the simplest order parameter is the average density of active sites $\rho(t)=n/N$. Without dilution ($p=0$) at the DP critical point $\lambda=\lambda_c$, this order parameter is known to decay algebraically as $\rho(t) \sim t^{-\delta}$ on an infinite lattice, where $\delta=\beta/\nu_\parallel\simeq 0.4505$ is the corresponding critical exponent. In what follows we study the diluted contact process along the line $p=1/2$, where the frozen network of bonds is at the percolation threshold. 

\subsection{Steady-state properties}
\label{StatSubSect}

The static exponent $\beta$ can be estimated by measuring the stationary density of active sites close to the transition. However, keeping $p=1/2$ fixed and varying $\lambda$ one finds that the stationary density of active sites $\rho^{stat}$ is always zero, even on line (b), where the contact process is locally supercritical. This is because the only surviving subprocess at the percolation threshold is the one running on the infinite percolation cluster while all other subprocesses on \textit{finite} clusters have a finite (although sometimes very long) life time. As the infinite percolation cluster with $D_f<2$ covers only a vanishing fraction of the lattice, this surviving subprocess does not contribute to the overall density of active sites. 

\begin{figure}[t]
\begin{center}
\includegraphics[width=105mm]{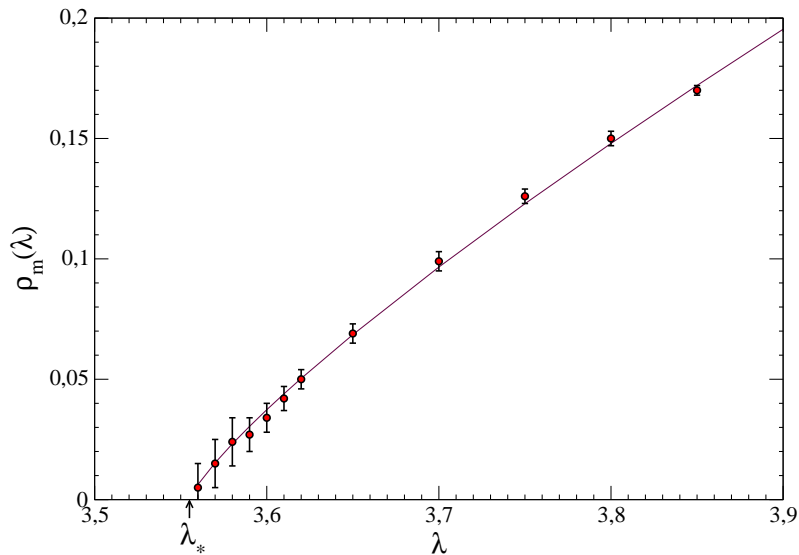}
\caption{
\label{fig:stationary}
Stationary density of active sites on the maximal cluster at the percolation threshold $p=p_c$ for various $\lambda$. The solid line shows a fitted power law (see text).
}
\end{center}
\end{figure}

In numerical simulations, however, one finds a quite stable positive density of active sites. The reason is that in finite system the maximal cluster still covers a considerable finite fraction of the lattice and therefore contributes to the overall density of active sites.  As the fractal dimension $D_f=91/48 \simeq 1.896$ of critical isotropic percolation is close to $2$ these finite-size corrections are enormous, e.g. on a lattice of $2000 \times 2000$ sites the largest cluster typically occupies about $30\%$ of the lattice. 

To circumvent this problem, we study the density of active sites on the \textit{maximal} cluster (which in a simulation approximates the infinite one). To this end we first dilute the lattice by removing bonds with probability $p$, classify all clusters using of the Hoshen-Kopelman algorithm~\cite{HoshenKopelman76}, identify the largest cluster and monitor the contact process exclusively on this cluster during the subsequent temporal evolution. The corresponding density, defined as the number of active sites on this cluster divided by its mass, will be denoted as $\rho_m(\Delta,t)$. 

Fig.~\ref{fig:stationary} shows the stationary density on the maximal cluster $\rho^{stat}_m$ as a function of $\lambda$ measured in a system of size $2000 \times 2000$ averaged over $100$ runs. Although the errors of the estimates are quite large, especially close to the critical point, the results are compatible with power-law scaling of the form
\begin{equation}
\rho_m \approx A (\lambda-\lamstar)^\beta\,,
\end{equation}
with
\begin{equation}
\lamstar=3.55(1) \,, \qquad \beta=0.81(7) 
\end{equation} 
and the proportionality factor $A\approx 0.464$. 

\subsection{Temporal decay at the multicritical point}
\label{MCPSubSect}

\begin{figure}[t]
\begin{center}
\includegraphics[width=120mm]{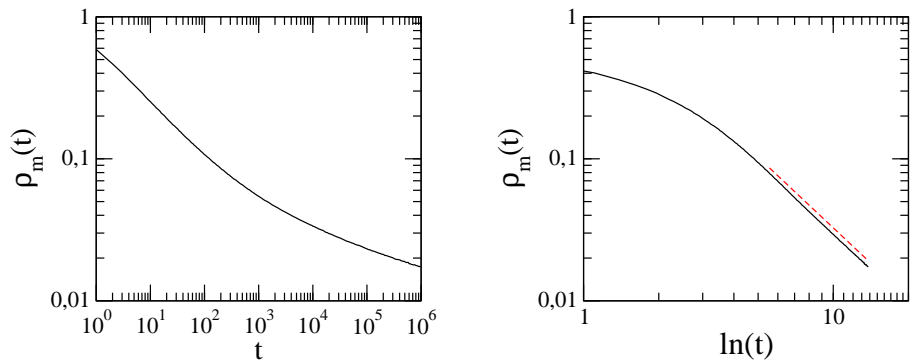}
\caption{
\label{fig:decay}
Decay of the density of active sites on the maximal cluster $\rho_m(t)$ at the multicritical point $p=p_c=1/2$ and $\lambda=\lamstar=0.355$ versus $t$ and $\ln t$ in a double-logarithmic representation. The dashed line indicates the slope -1.63.
}
\end{center}
\end{figure}

Having determined $\lamstar$ we study the temporal decay of $\rho_m(t)$ at the multicritical point. The numerically determined curve is plotted in Fig.~\ref{fig:decay} in two different representations. The left panel shows the data in an ordinary double-logarithmic plot. As can be seen, the line has an overall curvature, hence conventional power-law scaling is very unlikely. However, if the same curve is plotted against $\ln(t)$ in a double logarithmic plot, one obtains after an initial transient a straight line extending over almost four decades in time, as shown in the right panel of Fig.~\ref{fig:decay}. This suggests a logarithmically slow decay of the form
\begin{equation}
\label{LogarithmicDecay}
\rho_m(t) \sim [\ln(t/t_0)]^{-\deltabar}
\end{equation} 
with an exponent $\deltabar = 1.63(10)$ and the constant $t_0=1$. We note that the choice of $t_0$ is crucial for the estimate of the exponent (see discussion in Sect.~\ref{ConclusionSect}). In the present analysis we used $t_0=1$ for which the cleanest logarithmic decay is observed.

\subsection{Off-critical scaling}
\label{OffSubSect}

\begin{figure}[t]
\begin{center}
\includegraphics[width=140mm]{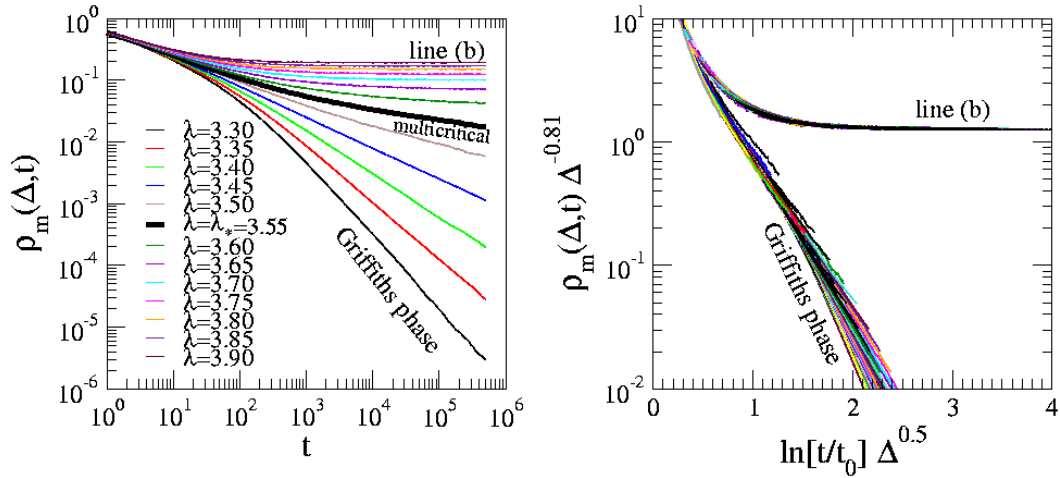}
\caption{
\label{fig:offcritical}
Off-critical simulations close to the multicritical point at the percolation threshold $p=1/2$. Left: Temporal decay of the density of active sites measured on the maximal cluster in a system with $2000 \times 2000$ sites for various values of the birth rate $\lambda$. The curve for $\lambda=\lamstar$, where the density decays logarithmically with time, is marked by a bold line. Right: Data collapse according to the scaling form~(\ref{OffScaling}) for $\lambda$ varying from $3.3$ to $3.9$ in steps of $0.05$. In order to avoid non-universal initial transients, this collapse shows only data points after $10$ Monte-Carlo sweeps.
}
\end{center}
\end{figure}

Next we study the temporal decay of $\rho_m(\Delta,t)$ for $\Delta=\lambda-\lamstar\neq 0$ in the vicinity of the multicritical point, keeping $p=1/2$ fixed. Our numerical results are shown in the left panel of Fig.~\ref{fig:offcritical}. As can be seen, for $\lambda<\lamstar$ ($\Delta<0$) one has straight lines with varying slopes, i.e. the system crosses over to an algebraic decay with continuously varying exponents which is the typical behavior in a Griffiths phase. For $\lambda>\lamstar$ ($\Delta>0$), i.e. on line (b) of the phase diagram, the contact process on the maximal cluster is supercritical so that the curves saturate at a constant value $\rho^{stat}_m(\Delta)$. 

Since the stationary density was shown to increase with $\Delta$ as a power law while at criticality the decay is logarithmically slow, it is very likely that the order parameter $\rho_m(\Delta,t)$ exhibits activated scaling. The corresponding scaling form can be derived from Eq.~(\ref{ActivatedScaling}) by inserting $\Lambda=\Delta^\nu_\perp$ and taking $L$ to infinity, giving
\begin{equation}
\label{OffScaling}
\rho_m(\Delta,t) = \Delta^\beta \, F\Bigl(\Delta^{\nuparbar}\,\ln(t/t_0)\Bigr)\,.
\end{equation} 
Hence, plotting $\rho_m(\Delta,t) \Delta^{-\beta}$ versus $\Delta^{\nuparbar}\,\ln(t/t_0)$ this scaling form can be tested by a data collapse. As shown in the right panel of Fig.~\ref{fig:offcritical}, the supercritical curves along line~(b) can be collapsed convincingly for $\nuparbar=0.50(5)$. The subcritical curves in the Griffiths phase, however, do not collapse entirely. This may be related to the non-universality of the power-law decay in the Griffiths phase.

\subsection{Finite-size scaling}
\label{FSSubSect}

So far we analyzed the scaling properties of the density of active sites $\rho_m(\Delta,t)$ on the maximal cluster, which in a numerical simulation can be regarded as approximating the infinite cluster at the percolation threshold. Let us now turn to the remaining clusters. Since in two-dimensional critical percolation the cluster sizes are distributed as~\cite{StaufferAharony92}
\begin{equation}
\label{ClusterSizeDistribution}
P(s) \sim s^{-\tau}=s^{-187/91}
\end{equation}
most of these clusters are very small so that finite-size effects are expected to play an important role. 

\begin{figure}[t]
\begin{center}
\includegraphics[width=91mm]{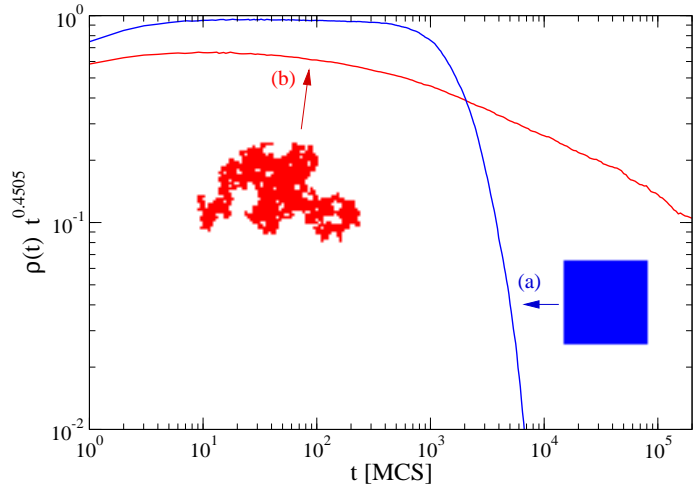}
\caption{
\label{fig:demo}
Decay of the density of active sites for (a) an contact process on a finite lattice with $N=32^2=1024$ sites at $\lambda=\lambda_c^{\rm \scriptscriptstyle 2D}$ averaged over $27000$ runs compared to (b) a contact process on a percolation cluster with approximately $1024$ sites at $\lambda=\lamstar$, averaging over $35000$ independent clusters. In both cases the density $\rho(t)$ is multiplied by $t^{\delta}$.}
\end{center}
\end{figure}

As illustrated in Fig.~\ref{fig:demo}, finite-size effects on finite percolation clusters differ significantly from ordinary finite-size effects. As an example the figure shows the decay of the density of active sites for (a) a critical contact process on a finite lattice with $32^2=1024$ sites, compared to (b) a contact process running on a percolation cluster with approximately $1024$ sites, averaged over many independent runs. As can be seen, the resulting curves differ significantly. Qualitatively this difference can be explained as follows:
\begin{itemize} 
\item[(a)] On a finite lattice the system first evolves in the same way as in an infinite system, exhibiting a clean power-law decay $\rho(t) \sim t^{-\delta}$ and a growing correlation length $\xi_\perp(t) \sim t^{1/z}$. As soon as $\xi_\perp(t)$ becomes comparable with the lateral size $L$ of the system (32 sites in Fig.~\ref{fig:demo}), the initial power law crosses over to an exponential decay. As can be seen in the figure, this finite-size effect sets in quite suddenly at a well-defined typical time. 

\item[(b)] On a percolation cluster with about the same number of sites, finite-size effects seem to influence the decay of the particle density at \textit{any} time $t$, leading to a soft overall curvature of the density in the double-logarithmic plot. This is plausible since on a percolation cluster there is no well-defined distance from boundary to boundary, instead the fringes of the cluster induce a whole spectrum of length scales which have to be compared with the growing correlation length of the contact process. Therefore, finite-size effects are always present and become gradually more important as time proceeds.
\end{itemize}
To analyze the decay of the particle density on finite clusters at the multicritical point, we first classify all clusters at the beginning of each run using the Hoshen-Kopelman algorithm~\cite{HoshenKopelman76}. For the subsequent statistical analysis the cluster sizes $s$ (i.e., the number of sites occupied by the clusters) are grouped logarithmically into bins with a density of 10 intervals per decade. For our simulations on a lattice with $2000 \times 2000$ sites this requires about 67 bins. For each bin we monitor the activity of all corresponding clusters as a function of time, averaging over $100$ independent runs. In this way we obtain a set of 67 curves parameterized by $s$, each of them describing the temporal evolution of the density of active sites on clusters with typical size $s$.

\begin{figure}[t]
\begin{center}
\includegraphics[width=140mm]{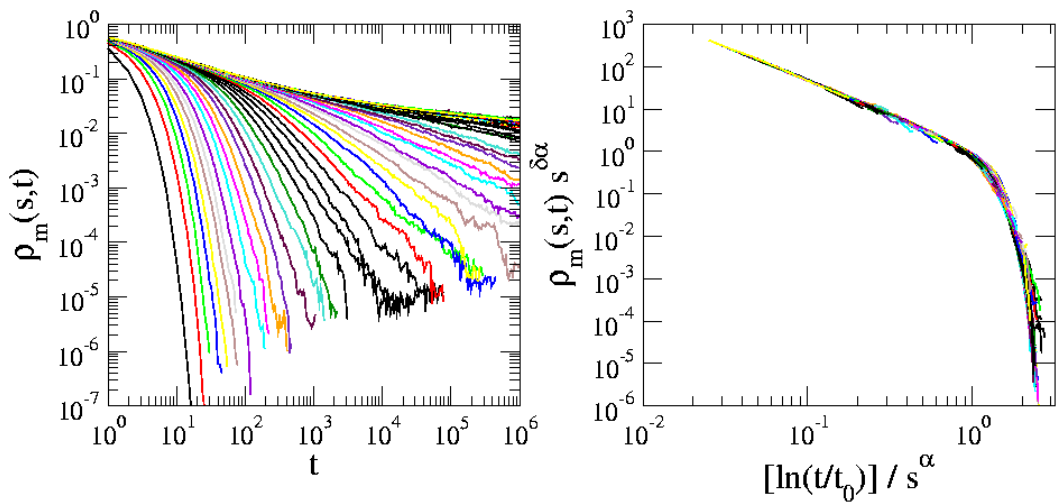}
\caption{
\label{fig:fs}
Finite size scaling analysis. Left: Density active sites for various cluster sizes $s$ ranging from 5 to $4 \times 10^6$ and grouped geometrically into 10 bins per decade. Right: Data collapse according to the scaling form (\ref{TestedFSScaling}), as explained in the text.
}
\end{center}
\end{figure}

In analogy to the off-critical case we test the possibility of activated scaling for finite clusters by a data collapse. Here the lateral system size $L$ in Eq.~(\ref{ActivatedScaling}) has to be replaced by the typical lateral length scale $\ell$ of a finite percolation cluster in two dimensions. For a cluster with $s$ sites this length scale is expected to be proportional to $\ell \sim s^{1/D_f}$, where $D_f=91/48 \approx 1.896$ is the fractal dimension of isotropic percolation in two dimensions. Inserting $\Lambda=\ell^{-1}=s^{-1/D_f}$ into Eq.~(\ref{ActivatedScaling}) one obtains the scaling form
\begin{equation}
\label{TestedFSScaling}
\rho(s,t)=s^{-x/D_f}\,F\Bigl(s^{-\zbar/D_f} \ln(t/t_0) \Bigr)\,.
\end{equation} 
%
%
Hence, plotting $\rho(s,t) s^{\beta/(\nu_\perp D_f)}$ versus $\ln(t/t_0)/s^{\zbar/D_f}$ this scaling form can be tested numerically by a data collapse. As shown in Fig.~\ref{fig:fs} a convincing data collapse is obtained for $\alpha=\zbar/D_f=0.30(2)$, hence we arrive at the estimate
\begin{equation}
\zbar=0.57(4)\,.
\end{equation} 
%
\subsection{Failure of logarithmic scaling}
\label{LogSubSect}
\begin{figure}[t]
\begin{center}
\includegraphics[width=100mm]{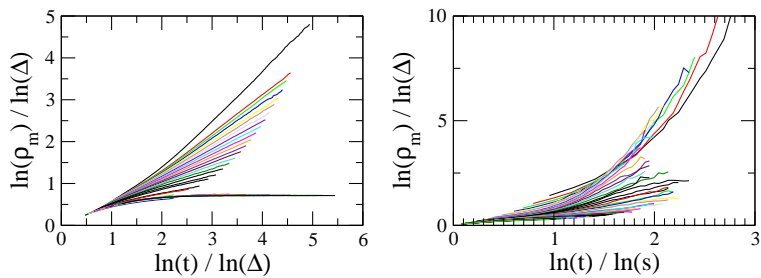}
\caption{
\label{fig:nolog}
Failure of logarithmic scaling according to Eq.~(\ref{AdaptedLogScaling}) (see text).
}
\end{center}
\end{figure}

%
Finally we demonstrate that generalized scaling with continuously varying exponents according to Ref.~\cite{SittlerHinrichsen02} seems to fail in the present model. As outline in Sect.~\ref{ContExpSubsect}, this theory predicts a logarithmic scaling form which in the present case reads:
\begin{eqnarray}
\label{AdaptedLogScaling}
\ln[\rho(\Delta,t,s)] &=& {\ln(s/s_0)} G \left( \frac{\ln(\Delta/\Delta_0)}{\ln(s/s_0)}, \frac{\ln(t/t_0)}{\ln(s/s_0)} \right) \\
&& +\, H \left( \frac{\ln(\Delta/\Delta_0)}{\ln(s/s_0)}, \frac{\ln(t/t_0)}{\ln(s/s_0)} \right) \nonumber\,.
\end{eqnarray}
Unlike activated scaling, this type of scaling requires four normalization constants $t_0, s_0$ and $\rho_0$ which may be regarded as fit parameters. Typically one expects these constants to be of the order 1 in natural units of the simulation. However, even when these constants are varied, we do not find convincing data collapses for both off-critical and finite-size simulations. For $H=0$ the failure of the data collapses is demonstrated in Fig.~\ref{fig:nolog}, where we set $\Delta_0=t_0=s_0=\rho_0=1$.

There is also a theoretical hint why this scaling form fails: Taking $\Delta\to 0$ and $s\to \infty$ one can show that Eq.~(\ref{AdaptedLogScaling}) reduces to $\frac{\ln[\rho(t)]}{\ln(t/t_0)} = const$, i.e., $\rho(t)$ has to decay algebraically provided that this constant is nonzero. As demonstrated in Sect.~\ref{MCPSubSect}, this is not the case, instead a logarithmic decay is found. 

\section{Discussion}
\label{ConclusionSect}

\begin{table}[t]
\begin{center}
\begin{footnotesize}
\begin{tabular}{|c|c|c|c|c|c|c|c|}
transition	&  quantity 	& scaling type	
&  $\beta$	&  $\nu_\perp$ 		&  $z$,$\zbar$	&  $\delta,\deltabar$ 	&  Ref. 		\\ \hline 
DP (2d)		& $\rho,\rho_m$	&  conventional 	
&  0.556(5) 	&  0.76(1) 		&  1.766(1) 	&  0.451(1)		&  \cite{LubeckReview} 	\\
line (a)	& $\rho,\rho_m$	&  activated 		
&  0.93(5) &  1.00(9) &  1.98(12) &  0.47 &  \cite{Dickman98}\\
line (b)	& $\rho$	&  activated 		
&  5/36 &  4/3 &  91/48 &  5/91 &  \cite{VojtaLee06} \\
MCP		& $\rho_m$	&  activated 		
&  0.81(7) &  0.88(10) &  0.57(4) &  1.63(10) &  this work
\end{tabular} 
\end{footnotesize}
\end{center}\caption{\label{table:exponents}
Numerical estimates of the critical exponents. } 
\end{table}

In the present paper we have analyzed the scaling behavior of the bond-diluted contact process at the percolation threshold $p=1/2$ for various birth rates $\lambda$ in the vicinity of the multicritical point $\lambda=\lamstar=0.355(1)$. Along this line the diluted lattice decomposes into a large number of clusters according to a scale-free distribution, each of them hosting an independent contact process. Performing numerical simulations we arrive at the following results:
\begin{enumerate}
\item The overall density $\rho$, which monitors the average activity of all contact processes, is characterized by a complex mixture of finite-size effects. While the infinite cluster does not contribute to $\rho$ in an infinite system, it leads to severe corrections in numerical simulations even even if one uses very large lattice sizes. Therefore, as a more appropriate order parameter, we studied the density of active sites restricted to the maximal cluster, $\rho_m(t)$, which is non-zero along line (b) and thus allows us to define a static exponent $\beta$.

\item The temporal decay of this order parameter at the multicritical point $\rho_m(t)\sim [\ln(t/t_0)]^{-\deltabar}$ is logarithmically slow, suggesting activated scaling at and in the vicinity of the multicritical point. This conjecture is supported by off-critical simulations and by an analysis of the density decay on finite-size clusters at criticality.

\item The estimated exponents (see Table~\ref{table:exponents}) differ from those on line (a) and line (b), suggesting a crossover phenomenon similar those observed at ordinary multicritical points with power-law scaling.

\item A fully logarithmic scaling form suggested in Ref.~\cite{SittlerHinrichsen02} seems to fail in the present model.
\end{enumerate}

\begin{figure}[t]
\begin{center}
\includegraphics[width=110mm]{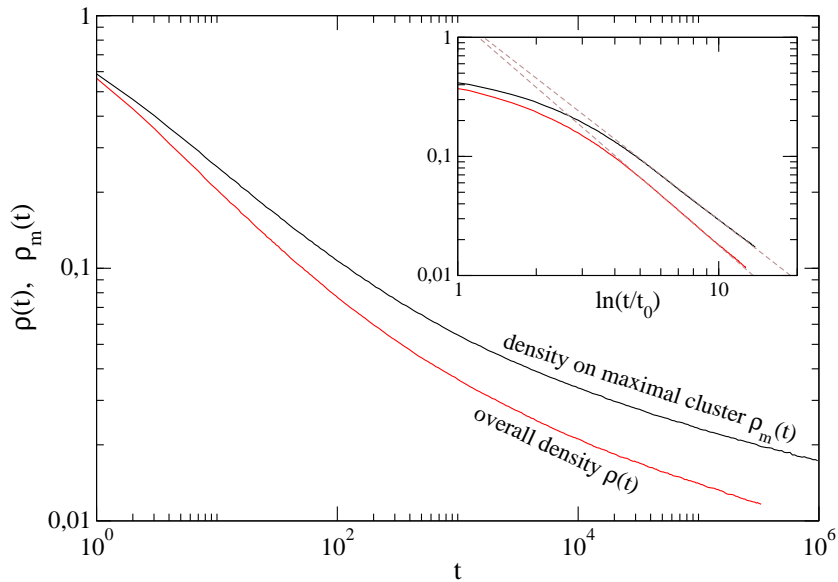}
\caption{
\label{fig:overall}
Overall density of active sites $\rho(t)$ compared to the density on the maximal cluster $\rho_m(t)$ versus $t$. As can be seen, $\rho(t)$ decays faster then $\rho_m(t)$. The inset shows the same curves versus $\ln t$ in a double-logarithmic representation. The slopes of the dashed lines indicate the theoretical prediction according to Eqs.~(\ref{LogarithmicDecay}) and~(\ref{Prediction}).
}
\end{center}
\end{figure}

As a cross-check we may now use these results in order to predict the decay of the overall density $\rho(t)$ at the multicritical point. To this end one has to integrate $\rho(s,t)$ over all cluster sizes, weighted by the cluster size $s$ and the probability $P(s) \sim s^{-\tau} = s^{-187/81}$ to find such clusters:
\begin{equation}
\rho(t) = \int {\rm d}s \, s \, P(s) \, [\ln(t/t_0)]^{-\deltabar} \, F\Bigl( [\ln(t/t_0)]/s^\alpha \Bigr)\,.
\end{equation} 
Dimensional analysis of this integral yields
\begin{equation}
\label{Prediction}
\rho(t) \sim [\ln(t/t_0)]^{-\deltabar - (2-\tau)/\alpha}\,.
\end{equation}  
This means that the overall density decays \textit{faster} than the density on the maximal cluster. In fact, this prediction can be verified numerically, as demonstrated in Fig.~\ref{fig:overall}.

Let us conclude with a critical remark. Unlike previous studies along lines (a) and (b), the results of the present work are based entirely on numerical evidence. Although the hypothesis of activated scaling at the multicritical point seems to be plausible, we realized that numerical simulations with an ultra-slow temporal evolution on a logarithmic time scale are very difficult to perform. In particular they are much more susceptible to misinterpretations caused by a wrong expectation. For example, if we expected power-law behavior at the transition, we would find an almost perfect-looking straight line in a log-log plot at $p=1/2$ and $\lambda=3.47$. Even data collapses in off-critical and finite-size simulations could be generated, but compared to the ones presented above their quality would be `less convincing'. The suggested scenario of activated scaling is much more coherent but still far from a proof.

Another important source of incertitude in scaling laws involving logarithms concerns the normalization constants needed to make arguments dimensionless. These constants are expected to be non-universal and have to be treated as fit parameters. For activated scaling the crucial constant is the time scale $t_0$. In our analysis we set $t_0=1$ for which the cleanest logarithmic decay of the density is seen. But obviously there is no particular reason for this particular choice, e.g. a simple redefinition of the rates in the Monte Carlo procedure would require a corresponding change of $t_0$. We were unable to specify a reliable error margin for $t_0$. For example, setting $t_0=1/2$ the data collapses would be slightly less convincing but the whole analysis still works but the estimates of the critical exponents seem to depend on the choice of $t_0$. This source of errors is not yet satisfactorily controlled. 


\vglue 2mm
{\bf \noindent Acknowledgments:}\\[1mm]
\noindent
This work was supported by the Deutsche Forschungsgemeinschaft (DFG) within the project WO-577/4.
S.R.D. would like to thank the Alexander von Humboldt Foundation for the financial support.

{\bf \noindent References:}

\end{document}